\documentclass{appolb}

\begin{document}
\pagestyle{plain}
\newcount\eLiNe\eLiNe=\inputlineno\advance\eLiNe by -1
\title{$NN\to NN\pi$ from the effective field theory point
of view: short comings and gains
\thanks{Contribution to "MESON2000", Workshop on production, properties and interaction of
 mesons, Cracow, Poland, 19-23 May 2000.}}
\author{C. Hanhart
\address{Department of Physics,  University of 
Washington, \\ Box 351560, Seattle, WA 98195-1560}}
\maketitle
\begin{abstract}
Based on a counting
 scheme for the application of chiral perturbation theory to
pion production in NN collisions p-wave production is studied.
It is demonstrated that, contrary to the s-wave production
where loops enter at too low an order to allow the scheme predictive
power, there are certain p-wave observables where parameter free predictions
are possible. These predictions are consistent with experimental data. 
We show that  the investigation of charged pion production yields
a  constraint on a contribution to the three-nucleon force
with a spin-isospin structure that could resolve the $A_y$ puzzle in $nd$
scattering. 
 
\end{abstract}

\newcommand{\boldpi}{\mbox{\boldmath $\pi$}}
\newcommand{\boldtau}{\mbox{\boldmath $\tau$}}
\newcommand{\boldT}{\mbox{\boldmath $T$}}
\newcommand{\gaprox}{$ {\raisebox{-.6ex}{{$\stackrel{\textstyle >}{\sim}$}}} $}
\newcommand{\saprox}{$ {\raisebox{-.6ex}{{$\stackrel{\textstyle <}{\sim}$}}} $}



In the last few years, the various $NN\rightarrow NN\pi$
reactions
have been studied both experimentally and theoretically
with a 
focus  on near-threshold energies. However, although
the first low energy data where published a decade
ago, the s-wave pion production is still not fully 
understood \cite{chrisreview}.

%
The pion dynamics are largely controlled by
chiral symmetry constraints, and the  hope that
the use  of Chiral Perturbation Theory ($\chi$PT) would yield
insights led to 
 the use of tree-level
$\chi$PT to calculate the cross sections close to threshold
\cite{CP2,CP1,CP3,HHH,CP4}.
The scale of typical internal momenta is set by the 
initial nucleon momentum $p_i \geq \sqrt{M_nm_\pi}$.
 Ref. \cite{CP2} emphasized that 
the diverse  contributions to the $Ss$ (capital letters denotes the $NN$ angular
momentum, whereas small letters denote the relative angular momentum of the
pion with respect to the $NN$ center of mass)
final states
thus have to be ordered
in powers of $\sqrt{m_\pi/M_N}$.
Because of this large scale 
the static approximation cannot be employed for nucleons.
More generally, as was pointed out in Ref. \cite{CP2,ulfrel},
there is a mix
among interactions of different chiral indexes.

\begin{figure}[tb]
\vspace{2.9cm}
\includegraphics{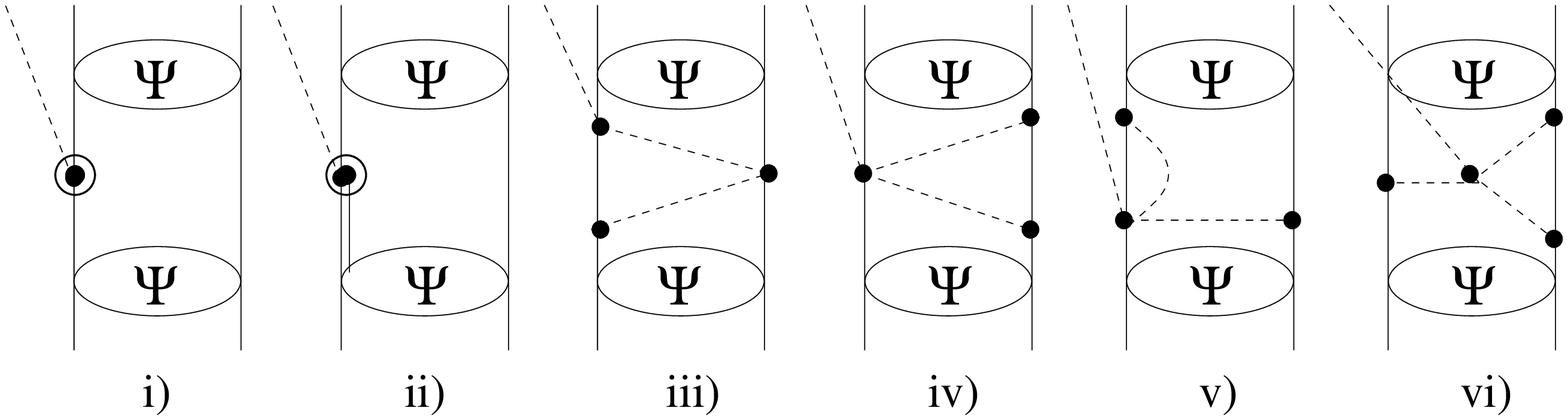}
\caption{ Leading (i \& ii) and 
some next-to-leading contributions to $s$-wave 
production.
A solid (dashed) line
denotes a nucleon (pion),
and a double line a $\Delta$.
Interactions from ${\cal L}^{(0)}$ (${\cal L}^{(1)}$)
are denoted by a dot (circled dot).
}
\label{pi0con} 
\end{figure}
Let us begin with a short discussion of s-wave production.
Due to a lack of space only $\pi^0$ production will  be looked
at here.
At leading order $({\cal O}(\sqrt{\frac{m_\pi}{M_N}}))$
direct emission off the nucleon and the delta contribute
(c.f. Fig. \ref{pi0con}i \& ii). Unfortunately the leading
order turns out to be small for two reasons that are not
under control of the power counting: i) a cancelation
of the nucleonic and delta diagram and ii) the presence of a 
zero in the half-off-shell NN T-matrix in the $^1S_0$ partial wave
at off-shell momenta $p_{off} \simeq p_i$ for small on shell momenta
\footnote{That is true for all
so called realistic potentials we are aware off.}.
At next-to leading order $({\cal O}(\sqrt{\frac{m_\pi}{M_N}}^2))$ already
loops enter. Some of those are depicted in figure \ref{pi0con}.
This explains the discrepancy of the 
results of \cite{CP1,CP2} and the data. 
A check of the convergence of the
series for the s-waves is thus technically difficult.

\begin{figure}[tb]
\vspace{2.9cm}
\includegraphics{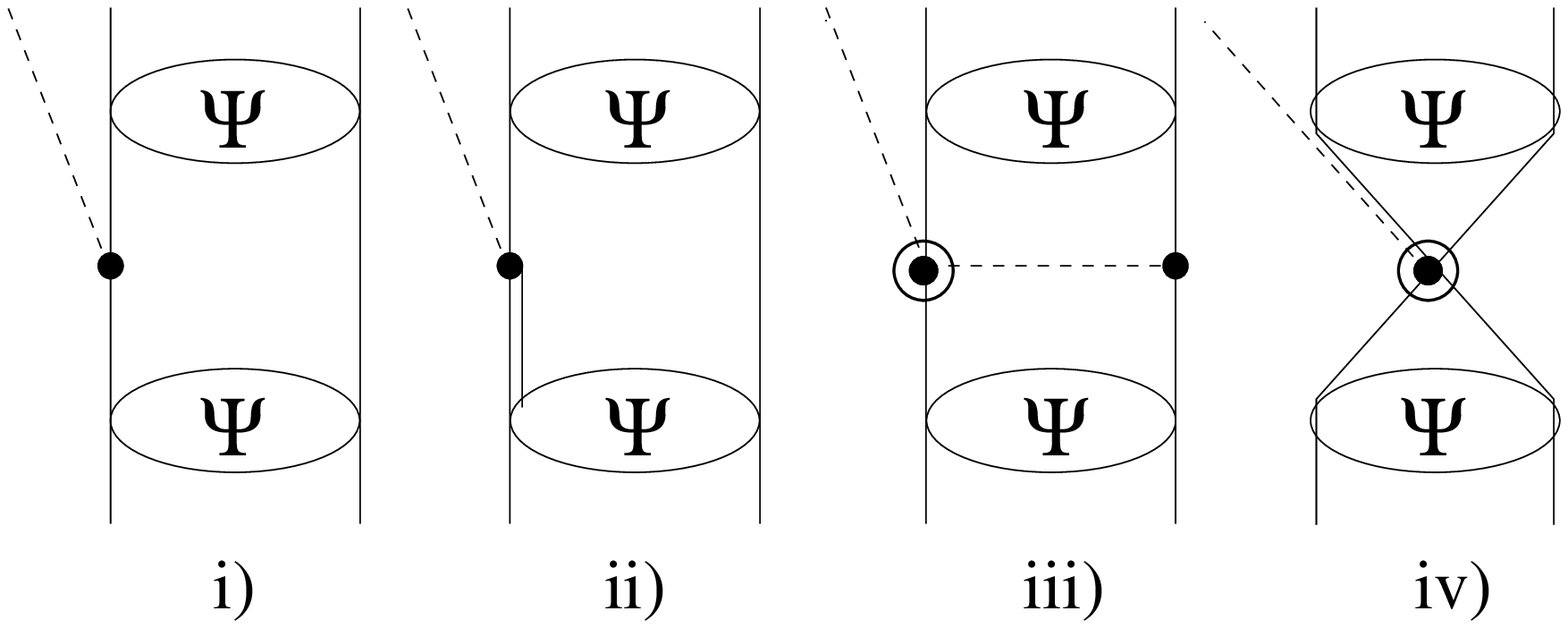}
\caption{ Lowest-order contributions to $p$-wave 
production.
Diagrams at 
${\cal O}(1)$ are (i,ii),
and of ${\cal O}(m_\pi/m_N)$ are (iii,iv). 
Diagrams with a 
$\Delta$ in the final state are also included. }
\label{pwave} 
\end{figure}

The situation is much more pleasant for the p-wave production as
was pointed out in ref. \cite{ours}. Let us
do the expansion at values of typical external momenta $q_{ex} = m_\pi$.
First of all the expansion starts at ${\cal O}(1)$, since the
leading operators lead to p-wave pions. Secondly, the next non 
vanishing order $({\cal O}(\sqrt{\frac{m_\pi}{M_N}}^2))$ is free of loops, since
for p-wave production not all the pion momenta in the loop can be internal ones.
Thus the order of the loops gets enhanced by a factor of $q_{ex}/p_{typ}=\frac{m_\pi}{\sqrt{m_\pi M_N}}$
compared to the s-wave case.
In figure \ref{pwave} we show the complete set of diagrams contributing to 
the p-wave $\pi^0$ production up to next-to-leading order. 
Thus we can employ chiral perturbation theory efficiently
in the study of pion production.

The Lagrangian relevant for the present investigation
is given in ref. \cite{ours}. As usual in effective field theories
a number of parameters appear whose strength is not constrained by
symmetry. In our case these are, besides the $\pi$-nucleon
coupling constant and $f_\pi$, the parameters in front of
terms with two pion fields and two nucleon fields
($c_3$ and $c_4$,
which where fixed in $\pi$N scattering \cite{BKM}) and two
further parameters in front of terms containing
four nucleon fields and one pion field -- $d_1$ and $d_2$. These are as yet unknown. 
It was pointed out in ref. \cite{bochum} that the latter structures can lead to
a significant contribution to the three nuclear force.
As will be shown below $\pi$--production in $NN$ collisions can
constrain the linear combination $d = d_1+4d_2$ and thus will
help us to get a consistent understanding of low energy nuclear
physics.

Before we move on, we first have to
show that the series converges, as claimed above. The crucial observation
to make this test possible is, that the short range $d_i$ terms only
support $S \to Sp$ transitions. We can therefore make 
parameter free predictions up to ${\cal O}(m_\pi/m_N)$ for all
those observables where the lowest $\pi$ partial wave allowed is p and not
both $NN$ states are in a relative S-wave. 
Such an observable exists, namely the $^3\sigma_1$ (we use the
notation of ref. \cite{meyer}: $^{2S+1}\sigma_m$) cross section
in neutral-pion production with $Pp$ as the lowest partial waves contributing.
This observable was recently measured at IUCF \cite{meyer}.
It was found in ref . \cite{ours} that indeed there is reasonable 
convergence in the p-wave production.

\begin{figure}[tb]
\vspace{5.2cm}
\includegraphics{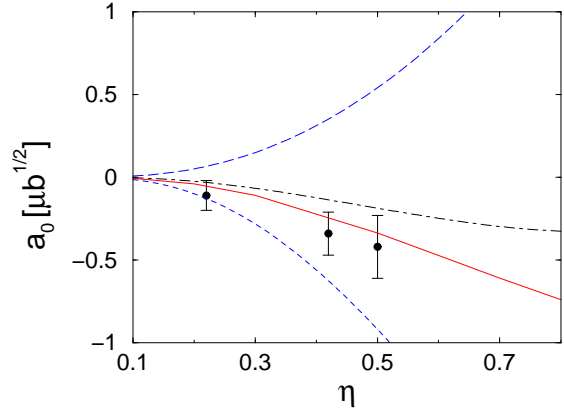}
\caption{ $a_0$
of  $pp \to np\pi^+$ in chiral perturbation theory.  
The different lines correspond to 
values of the parameter related to the three-nucleon force:
$\delta=1$ (long dashed line).
$\delta=0$  (dot-dashed line),
$\delta=-0.2$ (solid line), and
$\delta=-1$ (short dashed line).
Data are  from Ref. \protect\cite{Flammang}. }
\label{a0pn} 
\end{figure}

Thus we can move on to calculate an observable that is sensitive
to the short range operators proportional to $d$. Note that the
effective field theory not only constrains the possible coupling
structures but also gives an order of magnitude estimate for
the low energy constants. We can thus rewrite $d = \delta /(f_\pi^2M_N)$,
where we expect $\delta$ to be a number of order 1. The desired
observable is the amplitude of the $^1S_0\to ^3S_1p$ transition ($a_0$), recently extracted from
data of the reaction $pp\to pn\pi^+$ \cite{Flammang}. In Fig.
\ref{a0pn} we show the sensitivity of $a_0$ to the strength of the short range
interactions. Obviously, with better data pion production can put
a strong bound at a possible three body force originated in the
$d_i$ terms.

Based on the convergence of the chiral expansion in $p$-wave
pion production we demonstrated that data on this reaction can 
be used to extract information about the three-nucleon
force.
It is clear that more accurate data would be very useful.
We find it very gratifying that chiral symmetry
provides a direct connection
between pion production at energies $\sim 350$ MeV (IUCF)
and $Nd$ scattering at energies  
$\sim 10$ MeV (Madison, TUNL).


\begin{thebibliography}{10}

  
\bibitem{book} A.~M.~Bernstein and B.~R.~Holstein,
{\it  Chiral Dynamics: Theory and Experiment},
Springer (Berlin, 1995).


\bibitem{chrisreview}
C. Hanhart,  
STORI 99, Bloomington, Sept.  1999,
U.W. preprint  NT-UW-99-60,
{\tt nucl-th/9911023}. 

\bibitem{CP2b}
U.~van Kolck, G.A.~Miller, and D.O.~Riska,
Phys.\ Lett.\ {\bf B388} 679 (1996).

\bibitem{CP2}
T.D. Cohen, J.L. Friar, G.A. Miller, and U. van Kolck, 
Phys. Rev. C {53}, 2661 (1996); 
 U. van Kolck,   
B.A.P.S. {\bf 40}, 1629 (1995). 

\bibitem{CP1}
B.Y. Park et al., Phys. Rev. C {\bf 53}, 1519 (1996).

\bibitem{CP3}
T. Sato, T.-S.H. Lee, F. Myhrer, and K. Kubodera,
Phys. Rev. C {\bf 56}, 1246 (1997). 

\bibitem{HHH} C. Hanhart et al.,
Phys.\ Lett.\ {\bf B424}, 8 (1998).

\bibitem{CP4}
C. da Rocha, G. A. Miller, and U. van Kolck, 
Phys.\ Rev.\ C{\bf 61}, 034613 (2000).

\bibitem{ulfrel} V. Bernard, N. Kaiser and Ulf-G.~Mei\ss ner,
Eur. Phys. J A{\bf 4}(1999) 259.

\bibitem{ours}
C. Hanhart, G. A. Miller, and U. van Kolck, {\tt nucl-th/0004033}

\bibitem{BKM} V. Bernard, N. Kaiser, and Ulf-G. Mei{\ss}ner,
Nucl. Phys. {\bf B457}, 147 (1995).


\bibitem{bochum} 
D. H\"uber et al.,
Los Alamos preprint LA-UR-99-4996,
{\tt nucl-th/9910034}. 


\bibitem{meyer} H.O. Meyer et al., 
Phys.\ Rev.\ Lett.\  {\bf 83} (1999) 5439.

\bibitem{Flammang} R.W. Flammang et al., Phys. Rev. C {\bf 58}, 916 (1998).


\end{thebibliography}
\end{document}